\documentclass[journal]{IEEEtran}
\usepackage{multicol}
\usepackage{amsmath,amssymb,cite,multirow,subfigure}
\usepackage{graphicx}
\usepackage{url}
\usepackage[linesnumbered,ruled, vlined]{algorithm2e}
\usepackage{algorithmic}
\usepackage{makecell}

\usepackage{color}

\usepackage{booktabs}
\usepackage{threeparttable}

\title{An Efficient List Decoder Architecture for Polar Codes}
\author{Jun Lin and Zhiyuan Yan,~\IEEEmembership{Senior~Member,~IEEE}}
\begin{document}

\maketitle
\begin{abstract}
Long polar codes can achieve the symmetric capacity of arbitrary binary-input discrete memoryless channels under a low complexity successive cancelation (SC) decoding algorithm. However, for polar codes with short and moderate code length, the decoding performance of the SC algorithm is inferior. The cyclic redundancy check (CRC) aided successive cancelation list (SCL) decoding algorithm has better error performance than the SC algorithm for short or moderate polar codes. In this paper, we propose an efficient list decoder architecture for the CRC aided SCL algorithm, based on both algorithmic reformulations and architectural techniques. In particular, an area efficient message memory architecture is proposed to reduce the area of the proposed decoder architecture. An efficient path pruning unit suitable for large list size is also proposed.
For a polar code of length 1024 and rate $\frac{1}{2}$, when list size $L=2$ and 4, the proposed list decoder architecture is implemented under a TSMC 90nm CMOS technology. Compared with the list decoders in the literature, our decoder achieves 1.24 to 1.83 times hardware efficiency.
\end{abstract}
\section{Introduction}
\label{sec:intro}

Polar codes, recently introduced by Ar{\i}kan~\cite{arikan}, are a significant breakthrough in coding theory. It is proved that polar codes can achieve the channel capacity of binary-input symmetric memoryless channels~\cite{arikan} and the capacity of any discrete or continuous channel~\cite{sas_polar}. Polar codes can be efficiently decoded by the low-complexity successive cancelation (SC) decoding algorithm~\cite{arikan} with complexity of $O(N\log N)$, where $N$ is the block length.

Though the theoretical results are exciting, polar codes require very large code block length (for example, $N>2^{20}$~\cite{gross_polar1}) to approach the channel capacity using the SC algorithm. Such long block length is impractical in many applications, such as wireless communication systems where the packet size is only several hundred to several thousand bits. For short or moderate length, the error performance of polar codes with the SC algorithm is worse than Turbo or low-density parity-check (LDPC) codes~\cite{ido_list1}.

Lots of efforts~\cite{ido_list1, ido_list2, list2, list3, LP_polar, finite_length_polar, gcc_polar, kai_polar} have already been devoted to the improvement of error-correction performance of polar codes with short or moderate lengths. An SC list (SCL) decoding algorithm was proposed recently in~\cite{ido_list1}, which performs better than the SC algorithm and performs almost the same as a maximum-likelihood (ML) decoder~\cite{ido_list1}. In~\cite{ido_list2,list2, list3}, the cyclic redundancy check (CRC) is used to pick the output codeword from $L$ candidates, where $L$ is the list size. The CRC-aided SCL algorithm performs much better than the SCL algorithm at the expense of negligible loss in code rate. For example, it was shown~\cite{ido_list2} that the CRC-aided SCL algorithm outperforms the SC algorithm by more than 1 dB when the bit error rate (BER) is on the order of $10^{-5}$ for a polar code of length $N=2048$. The belief propagation (BP) algorithm on the factor graph of polar codes was investigated in~\cite{finite_length_polar}. It was shown~\cite{finite_length_polar} that finite-length polar codes show superior error floor performance compared to the conventional capacity-approaching coding techniques. In~\cite{gcc_polar}, polar codes were shown to be instances of generalized concatenated codes. It was suggested in~\cite{gcc_polar} that the performance of polar codes can be improved by considering them as generalized concatenated codes, and using block-wise near-maximum-likelihood decoding of optimized outer codes.

In terms of the hardware implementations of the SC algorithm, few works have been done. In~\cite{polar_fpga}, an FPGA implementation of a polar decoder based on belief propagation was proposed. An efficient semi-parallel SC decoder was proposed in~\cite{gross_polar1}, where resource sharing and semi-parallel processing were used to reduce the hardware complexity. An overlapped computation method and a pre-computation method were proposed in~\cite{chuan_polar} to improve the throughput and to reduce the decoding latency of SC decoders. Compared to the semi-parallel decoder architecture in~\cite{gross_polar1}, the pre-computation based decoder architecture~\cite{chuan_polar} can double the throughput. A simplified SC decoder for polar codes, proposed in~\cite{low_latency_polar}, reduces the decoding latency by more than 88\% for a rate 0.7 polar code with length $2^{18}$.

The investigation of efficient list decoder architectures for polar codes is motivated by improved error performance of the SCL and CA-SCL algorithms, especially for polar codes with short or moderate length. The tree search list decoder architecture for the SCL algorithm proposed in~\cite{tree_list_dec} is the only list decoder architectures for polar codes in the literature to the best of our knowledge. In this paper, we propose the first hardware implementation of the CA-SCL algorithm to the best of our knowledge. Based on both algorithmic and architectural improvements, our decoder architecture achieves better error performance and higher hardware efficiency compared with the decoder architecture in~\cite{tree_list_dec}. Specifically, the major contributions of this work are:
\begin{enumerate}
\item Message memories account for a significant fraction of an SC or SCL decoder~\cite{gross_polar1, tree_list_dec}.
In this paper, an area efficient message memory architecture is proposed. Besides, a new compression method for the channel messages is  used to reduce the area of the proposed decoder architecture.
\item An efficient processing unit (PU) is proposed. For the proposed list decoder architecture, a fine grained PU profiling (FPP) algorithm is proposed to determine the minimum quantization size of each input message for each PU so that there is no message overflow. By using the quantization size generated by the FPP algorithm for each PUs, the overall area of all PUs is reduced.
\item An efficient scalable path pruning unit (PPU) is proposed to control the copying of decoding paths. Based on the proposed memory architecture and the scalable PPU, our list decoder architecture is suitable for large list sizes.
\item A low-complexity direct selection scheme is proposed for the CA-SCL algorithm when a strong CRC is used (e.g. CRC32). The proposed direct selection scheme simplifies the selection of the final output codeword.
\item For a (1024, 512) rate-$\frac{1}{2}$ polar code, the proposed list decoder architecture is implemented for list size $L$ = 2 and 4, respectively, under a 90nm CMOS technology. Compared with the decoder architecture in~\cite{tree_list_dec} synthesized under the same technology, our decoder achieves 1.24 to 1.83 times hardware efficiency (throughput normalized by area). Besides, the proposed CA-SCL decoder has better error performance compared with the SCL decoder in~\cite{tree_list_dec}.
\end{enumerate}

The rest of this paper is organized as follows. In Section~\ref{sec:polar}, polar codes as well as the SCL and CA-SCL algorithms are briefly reviewed. Two improvements of the CA-SCL algorithm are discussed in Section~\ref{sec:fixed_point}. The proposed list decoder architecture is described in Section~\ref{sec:list_line}. Section~\ref{sec:result} shows the implementation and comparison results of the proposed list decoder architecture. The conclusions are drawn in Section~\ref{sec:conclusion}.

\section{Polar Codes and Its CA-SCL Algorithm}
\label{sec:polar}
\subsection{Polar Codes}
\label{ssec:polar_encoding}
A generation matrix of a polar code is an $N\times N$ matrix $G=B_NF^{\otimes n}$, where $N=2^n$, $B_N$ is the bit reversal permutation matrix~\cite{arikan}, and $F=\left[{1\atop 1}{0\atop 1}\right]$. Here $\otimes n$ denotes the $n$th Kronecker power and $F^{\otimes n} = F\otimes F^{\otimes (n-1)}$. Let $u_0^{N-1} = (u_0,u_1,\cdots,u_{N-1})$ denote the data bit sequence and $x_0^{N-1} = (x_0,x_1,\cdots,x_{N-1})$ the corresponding encoded bit sequence, then $x_0^{N-1}=u_0^{N-1}G$. The indices of the encoding bit sequence $u_0^{N-1}$ are divided into two sets: the information bits set $\mathcal{A}$ contains $K$ indices and the frozen bits set $\mathcal{A}^c$ contains $N-K$ indices. $u_{\mathcal{A}}$ are the information bits whose indices all come from $\mathcal{A}$. $u_{\mathcal{A}^c}$ are the frozen bits whose indices from $\mathcal{A}^c$.

\begin{algorithm}
\DontPrintSemicolon
\label{algo: scl}
\SetKwInOut{Input}{input}\SetKwInOut{Output}{output}

\Input{$n, \mbox{the received vector } y$}
\Output{$\hat{u}_0^{N-1}$}
\BlankLine
\For{$l=0$ \KwTo $L-1$} {
\For{$\beta =0$ \KwTo $N-1$}{
$P_{l,0}[\beta][s] = \Pr(y_{\beta}|s), s=0,1$\;
}
\lFor{$\lambda=0$ \KwTo $n$}{$r_l[\lambda] = 0$}
}

\For{$i =0$ \KwTo $N-1$}{
\lFor{$\lambda=\phi^i$ \KwTo $n-1$}{$r_l[\lambda] = l$}
\ForEach{survived decoding path $l$} {
metricComp$(l,i)$\;
}
\If{$i \in \mathcal{A}^c$} {
\ForEach{survived decoding path $l$} {
$\hat{u}_{l,i} = C_{l,n}[0][i \mod 2] = 0$\;
}
}
\Else {
pathPruning($P_{0,n},\cdots,P_{L-1,n}$)\; }
\If {$i \mod 2 == 1$} {
\ForEach{survived decoding path $l$} {
pUpdate$(l,n,i)$\;
}
}
}
\caption{SCL algorithm~\cite{ido_list1}}
\end{algorithm}

\begin{algorithm}
\DontPrintSemicolon
\label{algo: metricComp}
\SetKwInOut{Input}{input}\SetKwInOut{Output}{output}

\Input{$l, i$}
\BlankLine
determine $(b^{(i)}_{n},b^{(i)}_{n-1},\cdots,b^{(i)}_1)$ and $\phi^{(i)}$\;
\For{$\lambda = \phi^{(i)}$ \KwTo $n$} {
\For{$k=0$ \KwTo $2^{n-\lambda}$} {
\If{$b^{(i)}_{\lambda} = 1$ {\bf and} $\lambda = \phi^{(i)}$}{
$s = C_{l,\lambda}[\beta][0]$\;
$P_{l,\lambda}[k][u] $\\
$= G(P_{r_l[\lambda-1],\lambda-1}[2k],P_{r_l[\lambda-1],\lambda-1}[2k+1],s)$ \\
$= \frac{1}{2}P_{r_l[\lambda-1],\lambda-1}[2k][u\oplus s]\cdot P_{r_l[\lambda-1],\lambda-1}[2k+1][u] \mbox{ for } u\in\{0,1\}$ \\\;
}\Else{
$P_{l,\lambda}[k][u] $
$= F( P_{l,\lambda-1}[2k],P_{l,\lambda-1}[2k+1]) $\\
$= \sum\limits_{u'=0}^1\frac{1}{2}P_{l,\lambda-1}[2k][u\oplus u']\cdot P_{l,\lambda-1}[2k+1][u']$ \\
$\mbox{ for }u\in\{0,1\}$\;
}
}
}
\caption{metricComp$(l, i)$~\cite{ido_list1}}
\end{algorithm}

\subsection{SCL Algorithm}
List decoding was applied to the SC algorithm in~\cite{ido_list1} and the resulting SCL algorithm outperforms the SC algorithm. For a list size $L$, the SCL algorithm keeps at most $L$ decoding paths and outputs $L$ possible decoded codewords $\hat{u}_{0,0}^{N-1}, \hat{u}_{1,0}^{N-1}, \cdots, \hat{u}_{L-1,0}^{N-1}$, where $\hat{u}_{l,0}^{N-1} = (\hat{u}_{l,0}, \hat{u}_{l,0}, \cdots, \hat{u}_{l,N-1})$. A low complexity state copying scheme was proposed in~\cite{tree_list_dec} to simplify the copying process when a decoding path needs to be duplicated.

For $l = 0,1,\cdots, L-1$ and $\lambda=0,1, \cdots, n$, let $P_{l,\lambda}$ be an array with $2^{n-\lambda}$ elements: $P_{l,\lambda}[j]$ contains two messages $P_{l,\lambda}[j][0]$ and $P_{l,\lambda}[j][1]$ for $j=0,1,\cdots,2^{n-\lambda}-1$. $C_{l,\lambda}$ has the same structure as $P_{l,\lambda}$: $C_{l,\lambda}[j]$ contains two binary partial sums $C_{l,\lambda}[j][0]$ and $C_{l,\lambda}[j][1]$ for $j=0,1,\cdots,2^{n-\lambda}-1$. The SCL algorithm with low complexity state copying~\cite{tree_list_dec, ido_list1} is formulated in Algorithm~\ref{algo: scl}. For the decoding of $u_i$, the SCL algorithm can be divided into the following parts:
\begin{itemize}
\item For each surviving decoding path $l$, compute the path metrics $P_{l,n}[0][0]$ and $P_{l,n}[0][1]$ using the recursive function metricComp$(l,i)$ shown in Algorithm~\ref{algo: metricComp}. For $i=1,2,\cdots,N-1$, let ($b^{(i)}_{n},b^{(i)}_{n-1},\cdots,b^{(i)}_1$) denote the binary representation of index $i$, where $i=\sum_{j=0}^{n-1}2^jb^{(i)}_{n-j}$. $\phi^{(i)}$ $(1\leq \phi^{(i)} \leq n)$ in Algorithm~\ref{algo: metricComp} is the largest integer such that $b^{(i)}_{\phi^{(i)}}=1$. When $i=0$, $\phi^{(i)} = 1$. Based on the recursive algorithm for computing path metric in~\cite{ido_list1} and the low complexity state copying algorithm in~\cite{tree_list_dec}, the path metric computation is formulated in a non-recursive way in Algorithm~\ref{algo: metricComp}, where $\mathbf{r}_l=(r_l[n-1],r_l[n-2],\cdots,r_l[0])$ is the message updating reference index array for decoding path $l$. For decoding path $l$, $r_l[0]\equiv 0$, while all other elements are initialized with 0. Two types of basic operations, denoted as $F$ and $G$ operations, respectively, are employed in Algorithm~\ref{algo: metricComp}.
\item If $u_i$ is a frozen bit, for each decoding path, the decoded code bit $\hat{u}_{l,i} = 0$, decoding path $l$ will carry on with $\hat{u}_{l,i} = 0$.  If $u_i$ is an information bit, decoding path $l$ $(l=0,1,\cdots,L-1)$ splits into two decoding paths with corresponding path metrics being $P_{l,n}[0][0]$ and $P_{l,n}[0][1]$, respectively. There are at most $2L$ paths after splitting, and $2L$ associated path metrics. The pathPruning function in Algorithm~\ref{algo: scl} finds the $L$ most reliable decoding paths based on their corresponding path metrics.
\item For each of the $L$ surviving decoding paths, the pUpdate$(l,n,i)$ function shown in Algorithm~\ref{algo: pUpdate}~\cite{ido_list1} updates the partial sum matrices that will be used in the following path metric computation.
\end{itemize}

We make several observations about the path metric computation:
\begin{itemize}
\item When $i=0$, $P_{l,1},\cdots,P_{l,n}$ are updated in serial, and only the $F$ computation is employed.
\item For $i>0$, $P_{l,\phi^{(i)}},\cdots,P_{l,n}$ are updated in serial. The $G$ computation is used when computing $P_{l,\phi^{(i)}}$, while the $F$ computation is used for the other probability message arrays.
\item The computation of $P_{l,\phi^{(i)}}$ is based on $P_{r_l[\phi^{(i)}-1],\phi^{(i)}-1}$, while the computation of $P_{l,\lambda}$ ($\lambda > \phi^{(i)}$) is based on $P_{l,\lambda-1}$.
\end{itemize}
\begin{algorithm}
\DontPrintSemicolon
\label{algo: pUpdate}
\SetKwInOut{Input}{input}\SetKwInOut{Output}{output}

\Input{$l, \lambda, i$}
\BlankLine
\lIf{$\lambda == 0$} {\textbf{return}}
$j=\lfloor i/2\rfloor$\;
\For{$\beta =0$ \KwTo $2^{n-\lambda}-1$} {
$C_{l, \lambda-1}[2\beta][j \mod 2] = C_{l,\lambda}[\beta][0] \oplus C_{l,\lambda}[\beta][1]$\;
$C_{l, \lambda-1}[2\beta+1][j \mod 2] = C_{l,\lambda}[\beta][1]$\;
}
\lIf{$j \mod 2 == 1$} {pUpdate$(l,\lambda-1, j)$}
\caption{pUpdate$(l,\lambda, i)$~\cite{ido_list1}}
\end{algorithm}

\begin{figure*}
\begin{eqnarray}
~&G(P_{r_l[\lambda-1],\lambda-1}[2k],P_{r_l[\lambda-1],\lambda-1}[2k+1],s) =P_{r_l[\lambda-1],\lambda-1}[2k][u\oplus s]+P_{r_l[\lambda-1],\lambda-1}[2k+1][u] \label{equ: Gcomp_log}\\
~&F( P_{l,\lambda-1}[2k],P_{l,\lambda-1}[2k+1]) =\mbox{max}^*(P_{l,\lambda-1}[2k][u]+P_{l,\lambda-1}[2k+1][0], P_{l,\lambda-1}[2k][u\oplus 1]+P_{l,\lambda-1}[2k+1][1]) \label{equ: Fcomp_log}\\
~&F( P_{l,\lambda-1}[2k],P_{l,\lambda-1}[2k+1]) = \mbox{max}(P_{l,\lambda-1}[2k][u]+P_{l,\lambda-1}[2k+1][0], P_{l,\lambda-1}[2k][u\oplus 1]+P_{l,\lambda-1}[2k+1][1]) \label{equ: Fcomp_max}
\end{eqnarray}
\end{figure*}

The path pruning function, pathPruning, finds the $L$ most reliable paths, $a_0,a_1,\cdots,a_L$, and their corresponding decoded bits, $c_0,c_1,\cdots,c_L$, based on the path metrics. The path metrics of the surviving $L$ decoding paths are the $L$ largest ones among $2L$ input metrics. Once the surviving decoding paths are found, decoding path $l$ will copy from decoding path $a_l$. The partial sum computation of decoding path $l$ is carried on with the binary input $c_l$.

The pruning scheme in this paper and the path pruning scheme in~\cite{kai_polar} both try to eliminate decoding paths that are less reliable. However, there are still some differences as shown below.
\begin{itemize}
\item The pruning scheme in~\cite{kai_polar} is used for successive cancelation stack (SCS) decoding algorithm as well as the SCH decoding algorithm, which is a hybrid of SCL and SCS decoding algorithms, whereas our pruning scheme is used for the SCL algorithm.
\item For the SCL algorithm, suppose there are $L$ decoding paths before the decoding of $u_i$, then the metrics of $2L$ expanded decoding paths are computed. The pruning scheme in this paper finds the $L$ largest metrics out of $2L$ metrics and keeps their corresponding decoding paths. For the pruning scheme in~\cite{kai_polar}, a path will be deleted if its path metric is smaller than a dynamic threshold, $a_i - \ln(\tau)$, where $a_i$ is the largest metric of candidate paths, and $\tau$ is a configuration parameter.
\item For the path pruning scheme in~\cite{kai_polar}, the number of deleted paths is not fixed and depends on the configuration parameter $\tau$, while the number of deleted paths is always $L$ for the pruning scheme in this paper.
\end{itemize}

The SCL algorithm implemented in~\cite{ido_list1} is based on probability domain, where the $F$ and $G$ operations in Algorithm~\ref{algo: metricComp} are employed.
As shown in~\cite{list2}, the $F$ and $G$ operations in Algorithm~\ref{algo: metricComp} can also be performed over the logarithm domain. For $u\in \{0,1\}$, the resulting logarithm domain $G$ and $F$ computations are shown in Eq.~(\ref{equ: Gcomp_log}) and Eq.~(\ref{equ: Fcomp_log}), respectively, where $\max^*(x,y) = \max(x,y)+\log (1+e^{-|x-y|})$. $\max^*(x,y)$ can also be approximated with $\max(x,y)$, resulting in the approximated $F$ computation in~Eq.~(\ref{equ: Fcomp_max}).

\subsection{CA-SCL algorithm}
\label{ssec: ca_scl_algo}
In~\cite{ido_list2}, the performance of the SCL algorithm is further improved by the adoption of CRC, which helps to pick the right path from the $L$ possible decoded codewords.
In terms of the fixed point implementation, the CA-SCL algorithm is quite sensitive to saturation. For two decoding paths, it is hard to decide which is better if the metrics of both paths are saturated. In order to avoid message saturation, a non-uniform quantization scheme is proposed in~\cite{tree_list_dec}.
If the channel messages ($P_{l,0}$) are all quantized with $t$ bits, all the log-likelihood messages (LLMs) of $P_{l,\lambda}$ need to be quantized with $t+\lambda$ bits in order to avoid saturation. 

\section{Two Improvements of the CA-SCL Algorithm}
\label{sec:fixed_point}

\begin{figure} [hbt]
\centering
  \includegraphics[width=2in]{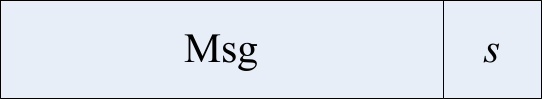}
  \caption{Compressed channel message}\label{fig: compressed_ch}
\end{figure}

In this paper, two improvements of the CA-SCL algorithm are proposed.
Firstly, for the $i$-th received bit $y_i$, there are two likelihoods, $\Pr \{y_i|0\}$ and $\Pr \{y_i|1\}$. Suppose $\Pr \{y_i|m\}$ $(m\in \{0,1\})$ is the smaller one among the two likelihoods. For $j\in\{0,1\}$, two log-likelihood messages (LLMs) are defined as
\begin{equation}
P_{l,0}[i][j] = \log\frac{\Pr \{y_i|0\}}{\Pr \{y_i|m\}}.
\end{equation}
Thus one of the LLMs is always 0, and the other is always non-negative. For the proposed list decoder, only the non-negative LLM and its corresponding binary index $s$ are stored. As shown in Fig.~\ref{fig: compressed_ch}, Msg denotes the stored non-negative LLM, and its corresponding bit index is $s$. When $s=0$, $P_{l,0}[i][0] =$ Msg, $P_{l,0}[i][1]=0$. When $s=1$, $P_{l,0}[i][0] = 0$, $P_{l,0}[i][1]=$ Msg. If $t$ bits are needed to quantize a channel LLM, it takes $t+1$ bits to represent two LLMs corresponding to a received bit $y_i$, while it takes $2t$ bits to store two LLMs directly.

Secondly, at the end of the CA-SCL decoding, the candidate codeword whose $K$ unfrozen bits pass the CRC is the output codeword. If more than one data word passes the CRC, it was proposed in~\cite{ido_list2} that the data word with the greatest path metric is chosen, which will incur additional comparisons. In this paper, a simple direct selection scheme is proposed: we first calculate all $L$ checksums in parallel and then scan from the checksum of data word 0 to the checksum of data word $L-1$, if a data word passes the CRC, the scan process is terminated and the corresponding candidate codeword is the final output one. When all $L$ CRC checks fail, since the CRC checksum could be corrupted, a decoding failure is announced if re-transmission is possible; otherwise, pick a data word randomly and output.

The direct selection scheme reduces computational complexity at the expense of possible performance degradation. In this paper, we give an estimation of the frame error rate (FER) degradation. Let $w$ denote the number of the detectable errors for our CRC. Assume all the bits of the final $L$ candidate data words are independently subject to a bit error probability, $p_b$. We calculate the increased FER, $P_e$, caused by the direct selection scheme instead of the ideal selection scheme, which always selects the transmitted data word if it is within the final $L$ candidates. For each candidate data word, there are three probabilities:
\begin{itemize}
\item The probability that the candidate data word is the same as the transmitted one is givn by $p_1 = (1-p_b)^K$.
\item The probability that the candidate fails the CRC is denoted as $p_2$.
\item The probability that the CRC identifies the candidate as the transmitted data word by mistake is denoted as $p_3$, and $p_3 \doteq \sum_{r=w+1}^{K}{K \choose r}p_b^{r}(1-p_b)^{K-r}\doteq {K \choose w+1}p_b^{w+1}(1-p_b)^{K-w-1}$.
\end{itemize}
Clearly, $p_1+p_2+p_3=1$.

Based on above assumptions and definitions, the increased FER
\begin{equation}
P_e \leqslant p_3\frac{1-p_2^L}{1-p_2}+p_2^L-(1-p_1)^L. \label{eqn: pe}
\end{equation}
Note that $p_b$ depends on the signal to noise ratio (SNR) and the list size $L$. For a specific SNR, in order to simplify our analysis, we can use $p_{b,\mbox{SC}}$ to approximate $p_b$, where $p_{b,\mbox{SC}}$ denotes the bit error probability of the SC algorithm. The probabilities, $p_2$ and $p_3$, are also approximated. Though approximated probabilities are employed when calculating $P_e$, the order of $P_e$ still helps us in determining whether our direct selection scheme is applicable. The impact of all the parameters are demonstrated in~(\ref{eqn: pe}). When a strong CRC is used, i.e. large $w$, $p_3$ is small, leading to a small $P_e$. On the other hand, a higher data rate leads to a greater $K$ and hence a greater $P_e$.

\subsection{Numerical Results}
\label{ssec:simulation}
\begin{figure} [hbt]
\centering
  \includegraphics[width=2.7in]{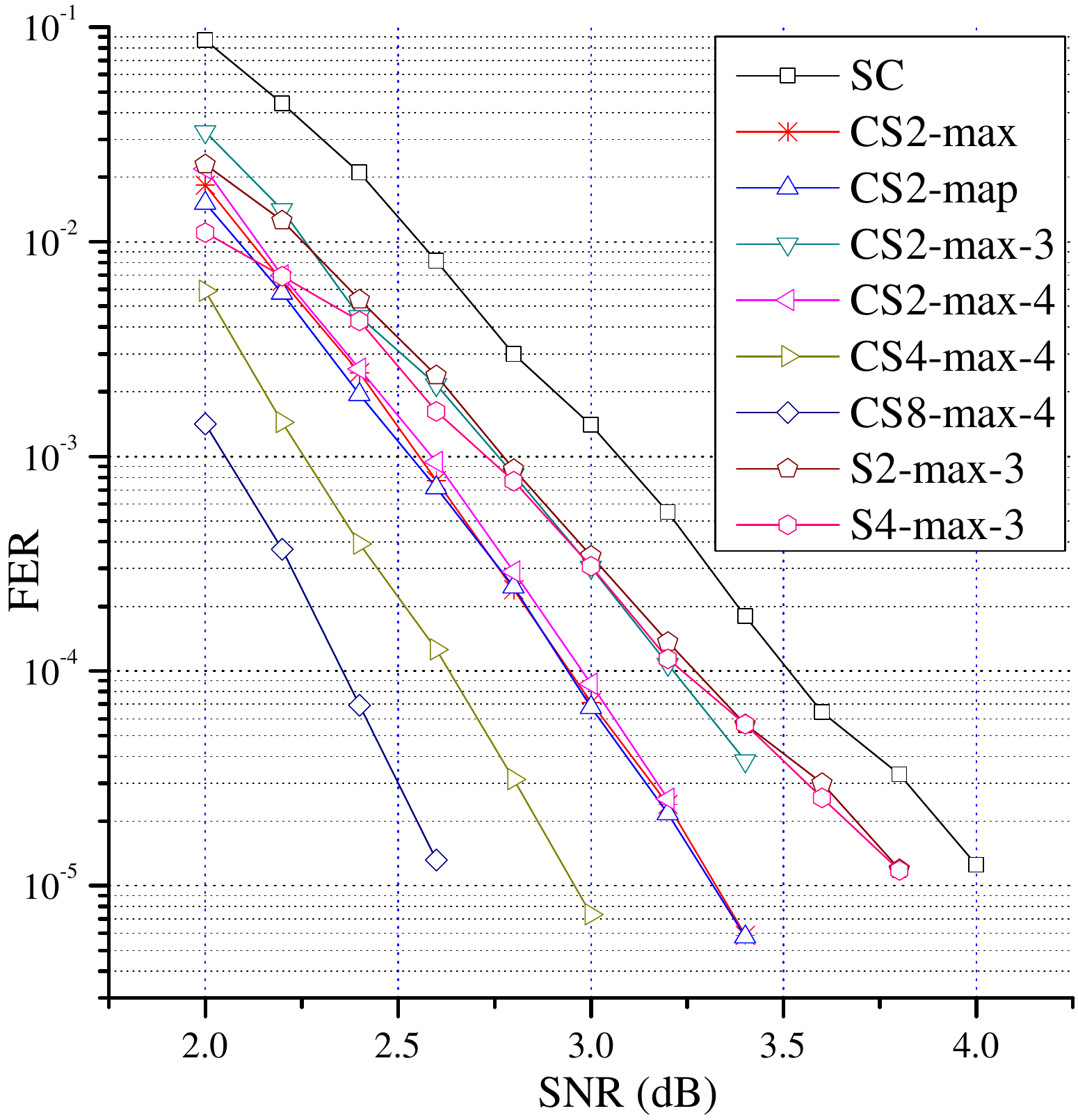}
  \caption{FER performance of a polar code with $N=1024$}\label{fig: fer}
\end{figure}

For a rate 1/2 polar code with block length $N=1024$, the frame error rate performances of the SC, SCL and CA-SCL algorithms are shown in Fig.~\ref{fig: fer}, where SC denotes the floating-point SC algorithm. CS2-max and CS2-map denote the floating-point CA-SCL algorithm with $L=2$ and the approximated $F$ computation shown in Eq.~(\ref{equ: Fcomp_max}) and the $F$ computation shown in Eq.~(\ref{equ: Fcomp_log}), respectively. CS$i$-max-$j$ denotes the fixed-point CA-SCL algorithm with $L=i$ and non-uniform quantization scheme with $t=j$, where $t$ is the number of quantization bits for channel probability message. S$i$-max-$j$ denotes the fixed-point SCL algorithm with $L=i$ and non-uniform quantization scheme with $t=j$. For all simulated CA-SCL algorithms, the CRC32 scheme is employed, and the direct selection scheme is employed to pick the final output codeword from $L$ possible candidates. The generation polynomial of the CRC32 is 0x1EDC6F41.

Based on the simulated results, several observations can be made:
\begin{itemize}
  \item For the CA-SCL algorithm, the approximated $F$ computation in Eq.~(\ref{equ: Fcomp_max}) results in negligible performance degradation.
  \item When each channel LLM is quantized with 4 bits, the employment of the proposed non-uniform quantization scheme leads to negligible performance degradation. When each channel LLM is quantized with 3 bits, the resulting FER performance is roughly 0.2dB worse than that using 4-bit quantization.
  \item Using a larger list size ($L>2$) leads to obvious decoding performance improvement for the CA-SCL algorithm, whereas the SCL algorithm with $L=2,4$ has nearly the same decoding performance, especially in the high SNR region. For polar codes with moderate block length (e.g. $N=2^{11},2^{12},2^{13}$), similar phenomena has been observed in~\cite{ido_list2}.
\end{itemize}

In this paper, more simulation results on the proposed direct selection scheme are provided. There are three selection schemes employed in our simulations.
\begin{itemize}
\item The proposed direct selection (DS) scheme, which outputs the first codeword that passes CRC.
\item Ideal selection (IS) scheme, which always outputs the correct codeword if it exists in the final list.
\item Metric based selection (MS) scheme~\cite{ido_list2}, which outputs the codeword that has the maximal path metric among all codewords that have passed CRC.
\end{itemize}

\begin{figure} [hbt]
\centering
  \includegraphics[width=2.8in]{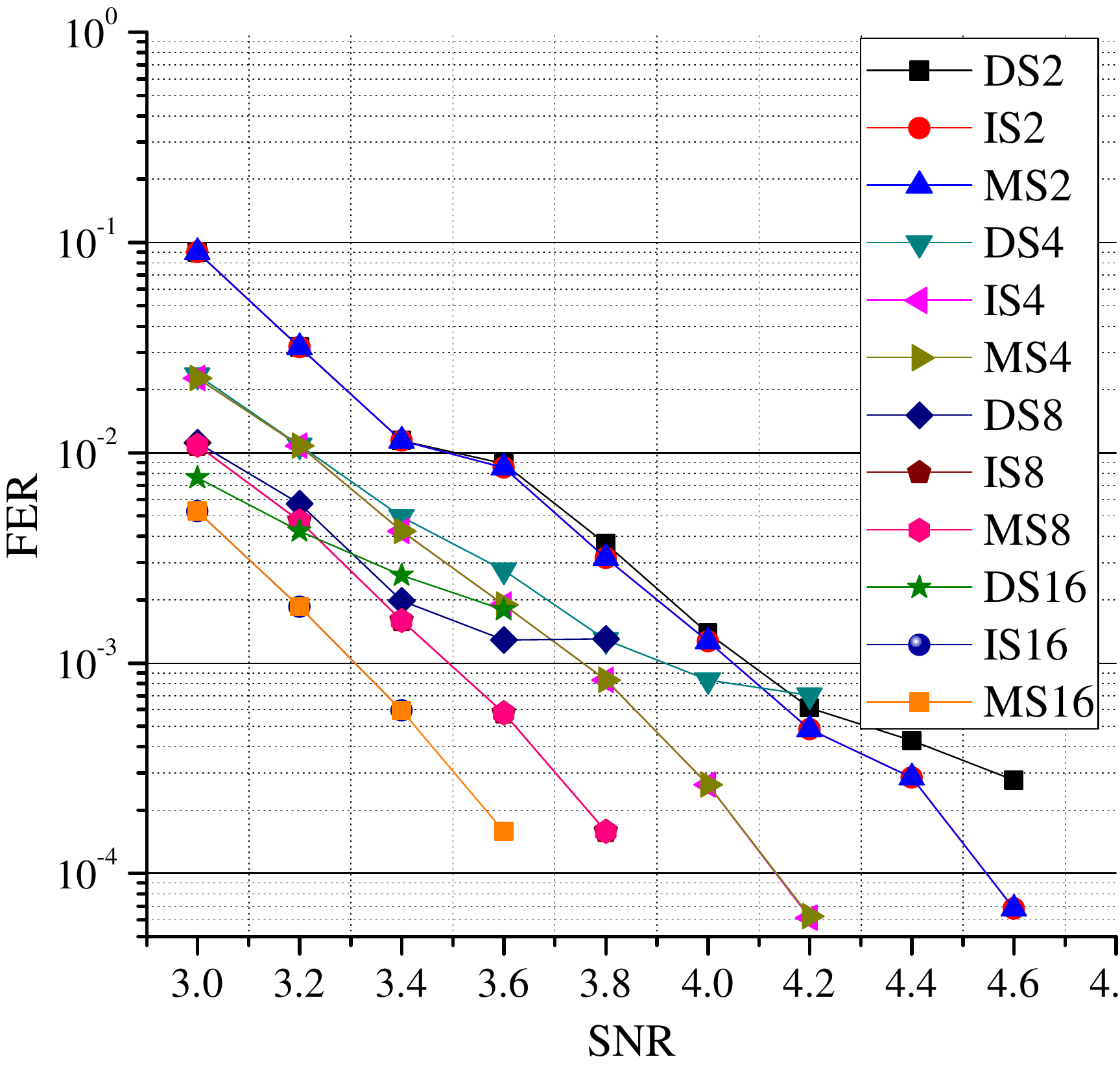}
  \caption{FER performances under CRC16 and rate 0.75}\label{fig: crc16r0_75}
\end{figure}

\begin{figure} [hbt]
\centering
  \includegraphics[width=2.8in]{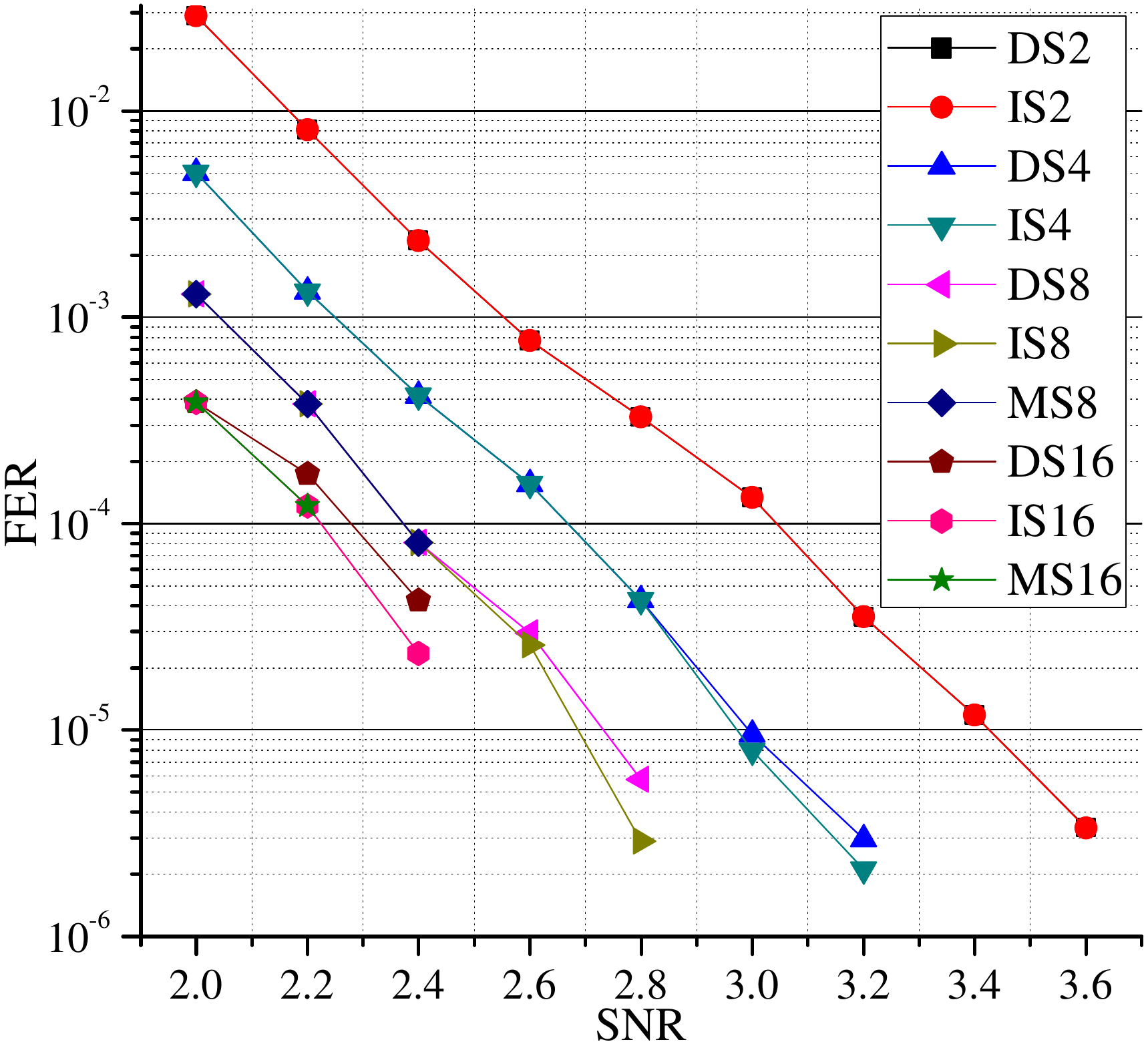}
  \caption{FER performances under CRC16 and rate 0.5}\label{fig: crc16r0_5}
\end{figure}

\begin{figure} [hbt]
\centering
  \includegraphics[width=2.8in]{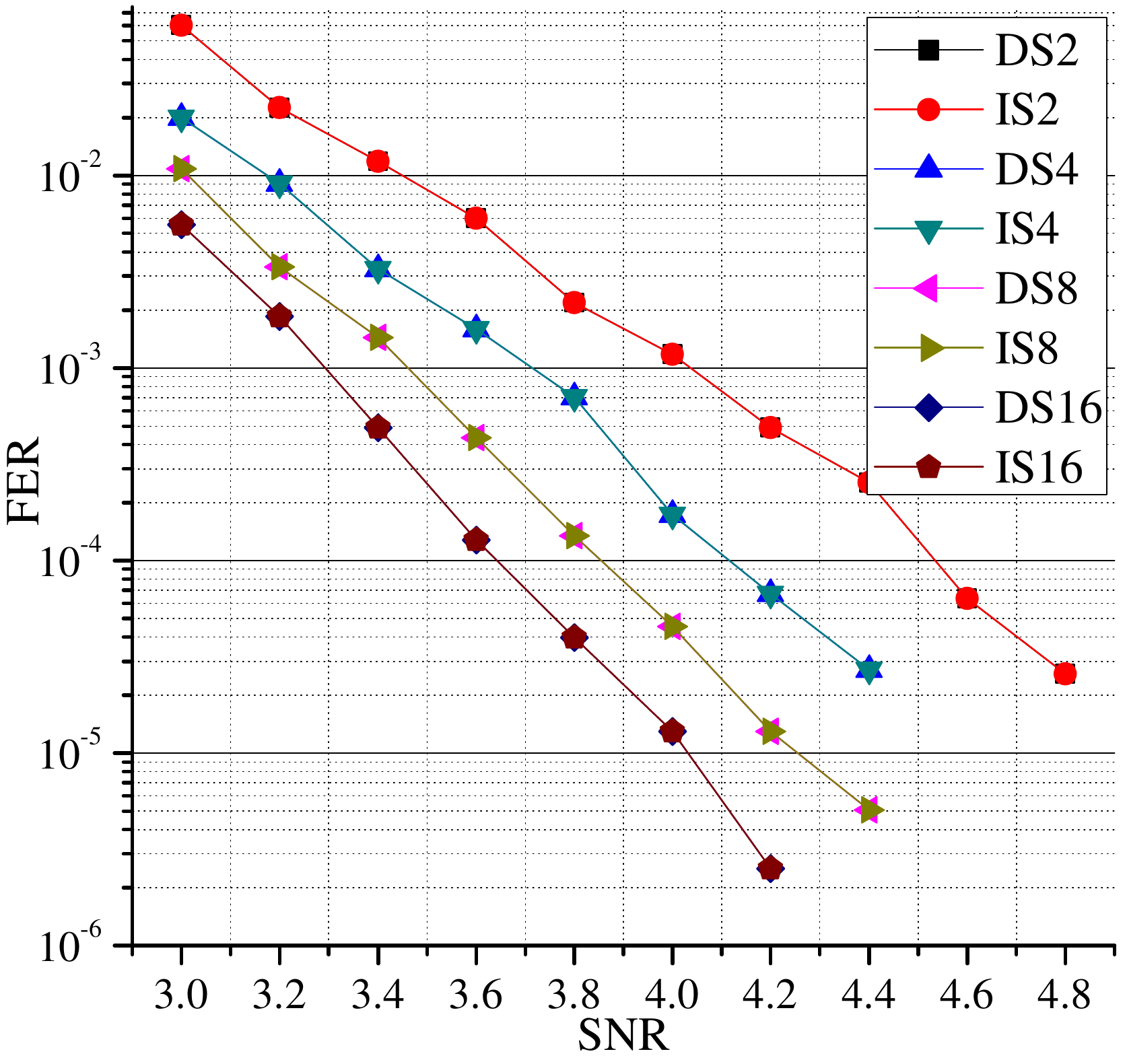}
  \caption{FER performances under CRC32 and rate 0.75}\label{fig: crc32r0_75}
\end{figure}

\begin{figure} [hbt]
\centering
  \includegraphics[width=2.7in]{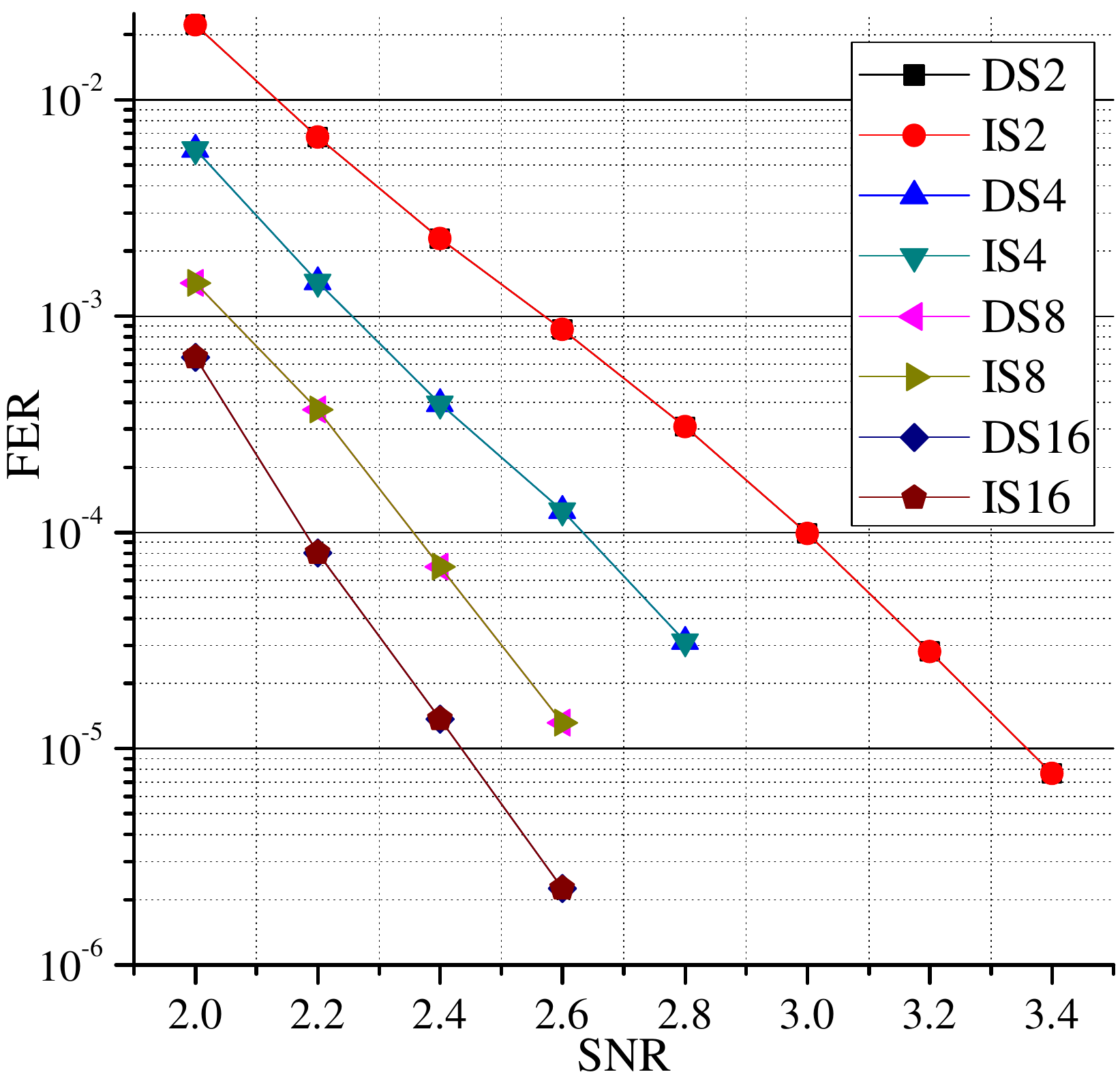}
  \caption{FER performances under CRC32 and rate 0.5}\label{fig: crc32r0_5}
\end{figure}

Still, the polar code of block length $N=1024$ is used in our simulations. In Figs.~\ref{fig: crc16r0_75} to~\ref{fig: crc32r0_5}, DS$k$, IS$k$ and MS$k$ denote the CA-SCL algorithms with list size $L=k$ under the direct selection scheme, the ideal selection scheme and the metric based selection scheme, respectively. The generation polynomial of the CRC16 used in our simulations is 0x1021.

As shown in Fig.~\ref{fig: crc16r0_75}, when code rate is 0.75, the proposed direct selection scheme introduces early error floor for all simulated list sizes while the metric based selection scheme performs nearly the same as the ideal selection scheme. When code rate is 0.5, as shown in Fig.~\ref{fig: crc16r0_5}, the direct selection scheme performs nearly the same as the ideal selection scheme with list size $L=2$. When list size $L=4,8,16$, the proposed direct selection scheme shows certain performance degradation compared with the ideal selection scheme, while the metric based selection scheme has little performance degradation. As shown in Figs.~\ref{fig: crc32r0_75} and~\ref{fig: crc32r0_5}, when CRC32 is used, the proposed direct selection scheme performs nearly the same as the ideal selection scheme for both code rates 0.5 and 0.75. 

We also calculate the $P_e$'s for all simulated cases in Figs.~\ref{fig: crc16r0_75} to~\ref{fig: crc32r0_5}. We choose SNR = 3.6dB, since DS4, DS8 and DS16 begin to show an error floor in Fig.~\ref{fig: crc16r0_75}. For the length 1024 polar code, the bit error probability $p_b$ from the SC algorithm is $6.28\times 10^{-4}$ and $3.04\times 10^{-6}$ for rate 0.75 and 0.5, respectively. The underlying channel is AWGN and the modulation is BPSK. For CRC16 and CRC32, $w$ = 2~\cite{crc16_detection} and 4~\cite{crc32_detection}, respectively. When CRC16 is used, for each simulated list size, the order of $P_e$ is around $10^{-2}$ and $10^{-10}$ for rate 0.75 and 0.5, respectively. When CRC32 is used, for each simulated list size, the order of $P_e$ is $10^{-4}$ and $10^{-17}$ for rate 0.75 and 0.5, respectively. As shown in Figs.~\ref{fig: crc16r0_75} to~\ref{fig: crc32r0_5}, it is found that the error degradation caused by our DS scheme is big when the corresponding $P_e$ is big (e.g. $10^{-2}$). On the other hand, when $P_e$ is quite small (e.g. $10^{-17}$), our DS scheme leads to little performance degradation.

Based on our calculation results, for a given CRC and code rate, $P_e$ increases with the list size $L$. This observation indicates that the potential performance degradation caused by the DS scheme will increase when $L$ increases. This is consistent with the simulation results shown in Figs.~\ref{fig: crc16r0_75} and~\ref{fig: crc16r0_5}.

\begin{figure*} [hbt]
\centering
  \includegraphics[width=4.2in]{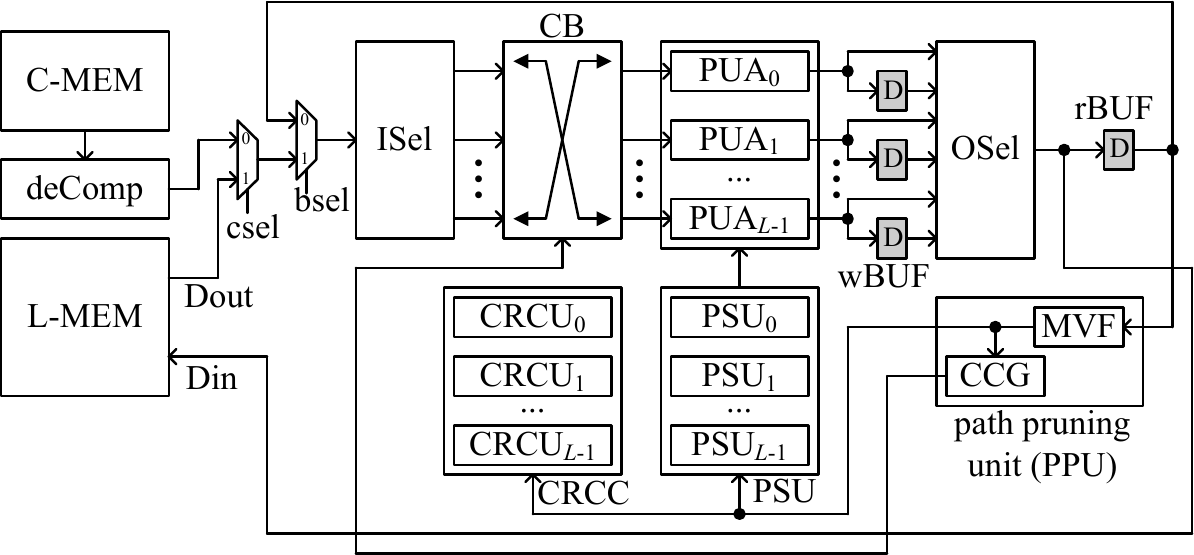}
  \caption{Top architecture of the list decoder}\label{fig: top_archi}
\end{figure*}

\section{Efficient List Decoder Architecture}
\label{sec:list_line}


For the CA-SCL algorithm, we propose an efficient partial parallel list decoder architecture shown in Fig.~\ref{fig: top_archi}. The proposed list decoder architecture mainly consists of the channel message memory (C-MEM), the internal LLM memory (L-MEM), $L$ processing unit arrays (PUAs) (PUA$_0$, PUA$_1$, $\cdots$, PUA$_{L-1}$), the path pruning unit (PPU) and the CRC checksum unit (CRCU). These components are described in details in the following subsections.

\subsection{Message Memory Architecture}
\label{ssec:l-mem-archi}

The L-MEM stores all the inner LLMs used for metric computation. Since all the LLMs in $P_{l,\lambda}$ need to be quantized with $t+\lambda$ bits for $\lambda \geq 1$, the variable-size LLMs make the L-MEM architecture for the proposed list decoder nontrivial. In this paper, an area efficient scalable memory architecture for L-MEM is proposed.
Due to the nonuniform quantization, the proposed L-MEM is built as follows.
\begin{itemize}
\item For $\lambda = 1,2,\cdots,n$, since each LLM within $\mathbf{P}_{\lambda} = (P_{0,\lambda}, P_{1,\lambda}, \cdots, P_{L-1,\lambda})$ is quantized with $t+\lambda$ bits, a regular sub-memory is created for storing LLMs in $\mathbf{P}_{\lambda}$.
\item All $n$ sub-memories are combined to a single memory.
\item Due to the nonuniform quantization, the width of each sub-memory maybe different. As a result, the concatenated L-MEM is an irregular memory with varying width within its address space. For the proposed memory architecture, the irregular L-MEM is split into several regular memories to fit current memory generation tools.
\end{itemize}
The proposed L-MEM is a mix of different types of memories, including SRAM, register file (RF) or register. Since SRAM and RF are more area efficient than a register, the proposed L-MEM architecture is better than the register based LLM memory in~\cite{tree_list_dec}\footnote{It was confirmed by the author of~\cite{tree_list_dec} that the LLM memory of the list decoder in~\cite{tree_list_dec} is built with registers.} especially for area restricting applications.

\begin{figure} [hbt]
\centering
  \includegraphics[width=2.6in]{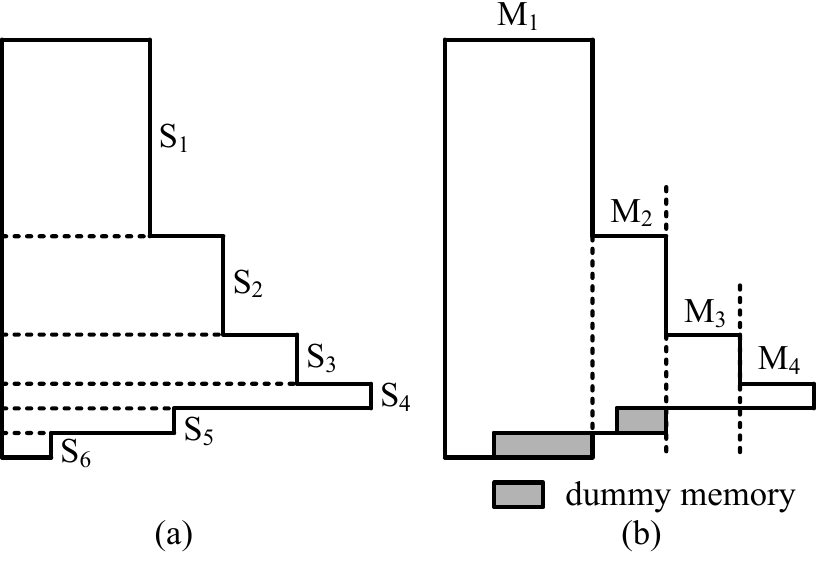}
  \caption{The split of an irregular LLM memory}\label{fig: mem_split}
\end{figure}

Suppose there are $T$ processing units (PUs) in each PUA shown in Fig.~\ref{fig: top_archi}, it consumes at most $4LT$ LLMs for one round of computation. For $\lambda = 1,2,\cdots,n$, we store all the LLMs within $\mathbf{P}_{\lambda} = (P_{0,\lambda}, P_{1,\lambda}, \cdots, P_{L-1,\lambda})$ in a single memory as follows.
\begin{itemize}
\item When $2^{n-\lambda+1}L > 4LT$, it takes a sub-memory of $\frac{2^{n-\lambda-1}}{T}$ words, where each word has $4LT(t+\lambda)$ bits.
\item When $2^{n-\lambda+1}L \leqslant 4LT$, it takes a sub-memory with only one single word, which has $2^{n-\lambda+1}(t+\lambda)L$ bits.
\end{itemize}
An example of the concatenation of $n=6$ sub-memories, (S$_1$, S$_2$, $\cdots$, S$_6$), is shown in Fig.~\ref{fig: mem_split}(a). For current memory compiler, it is hard to generate an irregular single memory instance as shown in Fig.~\ref{fig: mem_split}(a).

For the proposed L-MEM architecture, the concatenated irregular memory is split into several regular memory instances as shown in Fig.~\ref{fig: mem_split}(b), where additional dummy memories are added so that each instance is regular. For general cases, the irregular memory is divided into $\lambda_o = n-\log_2T-1$ regular instances. Depending on the number of words, each memory instance could be implemented with SRAM, RF or registers.

Compared with the register based LLM memory, the proposed L-MEM architecture is more area efficient due to the following reasons:
\begin{itemize}
\item Some sub-memory instances can be implemented with SRAM or RF which is more dense than register based memory.
\item As shown in Fig.~\ref{fig: mem_split}(b), most of the LLMs are store in the largest memory instance M$_1$ which contains
\begin{equation}
N_w = n-\lambda_o+1 +\sum_{\lambda = 1}^{\lambda_o-1}\frac{2^{n-\lambda-1}}{T} \label{eqn: mem_depth}
\end{equation}
words, where each word has $4LT(t+1)$ bits.
\end{itemize}

As shown in Eq.~(\ref{eqn: mem_depth}), $N_w$ is inverse to the number of processing units, $T$, within a PUA. As a result, the area of the proposed L-MEM depends on $T$ for a fixed block length $N=2^n$ and $t$. Taking RF as an example, we show the comparison of area efficiency of RFs with different depth in Table~\ref{tab:rf_size}, where area per bit (APB) denotes the total area of a memory normalized by the number of total bits. The total areas shown in Table~\ref{tab:rf_size} are from a memory compiler associated with a 90nm technology. As shown in Table~\ref{tab:rf_size}, the RF with a larger depth has a smaller APB. Hence, given the same amount of bits, it takes a smaller area if those bits can be stored in a RF with a larger depth. For SRAM, the same phenomena has been observed.

\begin{table}[hbt]
  \centering
  \caption{Area per Bit for RFs with Different Depth}
  \label{tab:rf_size}
  \footnotesize
  \begin{tabular}{c|c|c|c|c|c}
    \hline
     depth & 8 & 16 & 32 & 64 & 128  \\ \hline
     width & \multicolumn{5}{c}{128} \\ \hline
     process & \multicolumn{5}{c}{90nm} \\ \hline \hline
     total area ($\mu m^2$) & 24331 & 27022 & 32308 & 42812 & 63811 \\ \hline
     APB ($\mu m^2$) & 23.7 & 13.1 & 7.9 & 5.2 & 3.89 \\ \hline
  \end{tabular}
\end{table}

The C-MEM can be implemented with a simple regular memory, which has $\frac{N}{2T}$ words and each word has $2T(t+1)$ bits. Due to the proposed compression of the channel message, each compressed channel message is de-compressed into two LLMs before being fed to the PUs.

\subsection{Processing Unit Array}
\label{ssec: pua}

\subsubsection{Processing Unit Architecture}
\label{sssec: pu}
\begin{figure} [hbt]
\centering
  \includegraphics[width=2.0in]{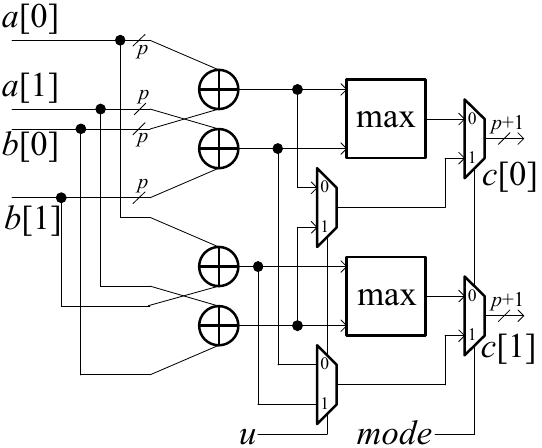}
  \caption{Processing unit (PU) architecture for the $G$ and approximated $F$ computations}\label{fig: pu}
\end{figure}

The $G$ and approximated $F$ computations shown in Eq.~(\ref{equ: Gcomp_log}) and Eq.~(\ref{equ: Fcomp_max}) , respectively, are used in the metric computation. These two types of basic operation can be performed with the PU shown in Fig.~\ref{fig: pu}, where $mode$ is the control signal and $u$ is the input partial sum for $G$ computation. The max module outputs the bigger value of the its two input values. When $mode = 0$, the approximated $F$ computation is performed. When $mode=1$, the $G$ computation is performed. These four adders in Fig.~\ref{fig: pu} are shared by both the $G$ and approximated $F$ computations. The hardware complexity of the proposed PU is determined by $p$, which is the width of an output LLM.

\subsubsection{Fine grained PU profiling}
\label{sssec: pu_profiling}
Due to the non-uniform quantization of the LLMs belonging to different message arrays, for each PU, the number of quantization bits, $p$, for each input LLM should be large enough so that no overflow will happen. According to the fixed point implementation of the CA-SCL algorithm, the quantization of $P_{l,n}$ ($l=0,1,\cdots,L-1$) needs the most binary bits, which is $t+n$. For each PUA, it is unnecessary to employ $T$ PUs with $p+1 = t+n$. In this paper, a fine grained PU profiling (FPP) algorithm, shown in Algorithm~\ref{algo: pu_profiling}, is proposed to decide $p$ for each PU.
\begin{algorithm}
\DontPrintSemicolon
\label{algo: pu_profiling}
\SetKwInOut{Input}{input}\SetKwInOut{Output}{output}
\Input{$n, t, \lambda_o = n-\log_2T-1$}
\Output{$p[0], p[1], \cdots, p[T-1]$}
\BlankLine
\For{$j=0$ \KwTo $T-1$}{
$p[j]$ = $t+\lambda_o-1$\;
}
\For{$\lambda = \lambda_o+1$ \KwTo $n$}{
\For{$j=0$ \KwTo $2^{n-\lambda}-1$}{
$p[j]$ = $t+\lambda-1$\;
}
}
\caption{FPP Algorithm}
\end{algorithm}

For the $j$-th PU of PUA$_l$ ($l=0,1,\cdots,L-1$), each LLM input is quantized with $p[j]$ bits. The proposed FPP algorithm is based on the observation that only $2^{n-\lambda} < T$ PUs are needed when computing the updated $P_{l,\lambda}$ with $\lambda > \lambda_o$. Thus, in the proposed PUA$_l$, only PU$_{l,0}$, PU$_{l,0}$, $\cdots$, PU$_{l,2^{n-\lambda}-1}$ are enabled for the computing of $P_{l,\lambda}$. Based on the proposed FPP algorithm, each PUA can finish the metric computation without any overflow at the cost of less area consumption.

As shown in Algorithm~\ref{algo: pu_profiling}, the bit width of the LLM inputs of a PU is determined by $n$, $T$ and $t$. One example is shown in Table~\ref{tab:pu_llm_width}, where $n=10$, $T=8$ and $t=4$.
\begin{table}[hbt]
  \centering
  \caption{Bit width of LLM Inputs of PU$_{l,j}$ when $n=10$, $T=8$ and $t=4$}
  \label{tab:pu_llm_width}
  \footnotesize
  \begin{tabular}{c|c|c|c|c|c|c|c|c}
    \hline
          $j$& 0 & 1 & 2 & 3 & 4 & 5 & 6 & 7  \\ \hline
      $p[j]$ & 13 & 12 & 11 & 11 & 10 & 10 & 10 & 10 \\ \hline
  \end{tabular}
\end{table}

The area saving due to the proposed fine grained profiling algorithm also depends on $T$, $n$ and $t$. For the proposed list decoder architecture, there are $L$ identical PU arrays, where each array contains $T$ PUs. In Table~\ref{tab:pu_area_cmp}, we compare the area of a regular PU array with that of an array where the input message width of each PU is determined by the fine grained profiling algorithm. As shown in Table~\ref{tab:pu_area_cmp}, the area of PU arrays is reduced by 30\% to 55\% depending on the number of PUs with an array and the block length $N=2^n$. Here, each channel message is quantized with $t=4$ bits.

\subsubsection{Metric Computation Schedule}
\label{sssec: metric_comp}
For the proposed L-MEM, each data word is capable of storing $4TL$ LLMs. Moreover, each word is equally divided into $L$ consecutive parts, where the $l$-th part stores the LLMs corresponding to decoding path $l$. The metric computation schedule is almost the same as that of the partial parallel SC decoder in~\cite{gross_polar1} except that $L$ PUAs work simultaneously for $L$ decoding paths, respectively.

When a data word needs to be updated, the write mismatch would happen since $L$ PUAs generate only $2LT$ updated LLMs during one clock cycle. These $L$ PUAs need to read two consecutive data words from L-MEM in order to generate $4TL$ LLMs. For the proposed list decoder architecture, as shown in Fig.~\ref{fig: top_archi}, $L$ write buffers (wBUFs) are employed to store half of $4TL$ LLMs generated by $L$ PUAs. Once the remaining LLMs are computed, the output selection (OSel) module formats these LLMs in the way that these LLMs are stored in the L-MEM.

Since all the LLMs belonging to $\mathbf{P}_{\lambda} = (P_{0,\lambda}, P_{1,\lambda}, \cdots, P_{L-1,\lambda})$ with $\lambda > \lambda_o$ are stored in a single data word in L-MEM and the computing of LLMs belonging to $\mathbf{P}_{\lambda+1}$ can only take place once $\mathbf{P}_{\lambda}$ are updated, an additional clock cycle is needed to read out the LLMs within $\mathbf{P}_{\lambda}$ that have been just written into the L-MEM. This will increase the delay and decrease the throughput of the proposed list decoder. As shown in~\cite{gross_polar1}, the bypass buffer, rBUF, is used to temporarily store the messages written into the L-MEM and eliminate the extra read cycle.

\begin{table*}[hbt]
  \centering
  \caption{Area comparison between fine grained PU array and regular PU array}
  \label{tab:pu_area_cmp}
  \footnotesize
  \begin{tabular}{c|c|c|c|c|c|c|c|c}
    \hline
     $n$ &\multicolumn{4}{c|}{10} & \multicolumn{4}{c}{15}\\ \hline\hline
     process & \multicolumn{8}{c}{TSMC 90nm CMOS} \\ \hline
     $T$             & 8    & 16    & 32     & 64    & 32      & 64    & 128   & 256 \\ \hline
     CPD (ns)        & \multicolumn{4}{c|}{0.555}& \multicolumn{4}{c}{0.588} \\ \hline
     regular PU array area ($\mu m^2$)& 27650 & 55259 & 113902 & 225418 & 150951 & 308640 & 602509 & 1212359 \\ \hline
     fine grained PU array area ($\mu m^2$)& 19280 & 34131 & 59048 & 101377& 104434 & 190615 & 334927 & 594048 \\ \hline
     area saving     & 30\% & 38\% & 48\% & 55\% & 30\% & 38\% & 44\% & 51\% \\ \hline
  \end{tabular}\\
\end{table*}

\subsection{Path Pruning Unit}
\label{ssec: path_pruning_unit}
For the CA-SCL algorithm, once the path metric computation of decoding step $i$ is finished, each current decoding path splits into two sub decoding paths. However, the list decoder keeps at most $L$ decoding paths. For the proposed list decoder architecture, a path pruning unit (PPU) is proposed to prune the split decoding paths in an efficient way. As shown in Fig.~\ref{fig: top_archi}, the proposed PPU contains two sub modules, the maximum value filter (MVF) and the crossbar control signals generator (CCG). The MVF generates $L$ path indices $a_0,a_1,\cdots,a_{L-1}$ and $L$ associated decoded bits $c_0,c_1,\cdots,c_{L-1}$. For a current decoding path $l$, both the path metric and partial sum computations will be based on the LLMs and partial sums within decoding path $a_l$, and the decoded code bit for $u_{l,i}$ is $c_l$. $a_l$ and $c_l$ for $l=0,1,\cdots,L-1$ are used to control the copying of partial sums and checksums.


\begin{table*}[hbt]
  \centering
  \caption{Comparison of ASIC implementation results}
  \label{tab:bitonic_cmp}
  \footnotesize
  \begin{tabular}{c|c|c|c|c|c|c|c|c|c|c}
    \hline
     &\multicolumn{5}{c|}{metric sorter~\cite{tree_list_dec}} & \multicolumn{5}{c}{proposed MVF}\\ \hline\hline
     process & \multicolumn{10}{c}{90nm CMOS} \\ \hline
     $L$             & 2    & 4    & 8     & 16    & 32      & 2    & 4    & 8 & 16 & 32 \\ \hline
     CPD (ns)        & 0.45 & 0.85 & 1.8   & 4.1   & 9.6     & 0.54 & 1.25 & 2.25 & 3.7 & 5.2\\ \hline
     area ($\mu m^2$)& 1995 & 9199 & 47119 & 241633& 1392617 & 1580 & 8401 & 30814 & 96979 & 319498\\ \hline
     area saving     &\multicolumn{5}{c|}{--}                & 20\%  & 8\% & 34\%  & 59\% & 77\% \\ \hline
  \end{tabular}\\
\end{table*}

\subsubsection{Maximum Values Filter}
\label{sssec: max_v_finder}

\begin{figure} [hbt]
\centering
  \includegraphics[width=3.2in]{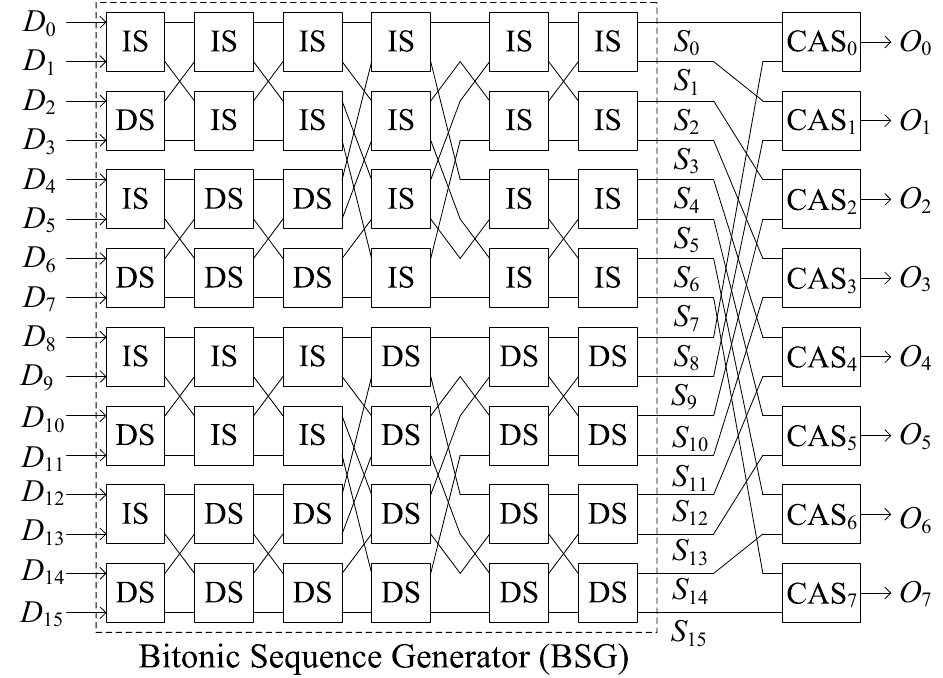}
  \caption{Maximum values filter architecture}\label{fig: mvf}
\end{figure}

Taking list size $L=8$ as an example, the corresponding MVF architecture is proposed in Fig.~\ref{fig: mvf}, where the MVF consists of a bitonic sequence generator (BSG) and a stage of compare and select (CAS) modules. The BSG has 16 inputs ($D_0, D_1,\cdots, D_{16}$) and 16 outputs ($S_0, S_1,\cdots, S_{16}$). Each of them consists of three parts: the path metric, the associated list index and decoded bit. The width of each input and output is $z=x_1+x_2+1$, where $x_1=t+n$ is the number of bit used to quantize a path metric and $x_2=\log_2L$ is the number of bits used to represent a list index.

\begin{figure} [hbt]
\centering
  \includegraphics[width=3.1in]{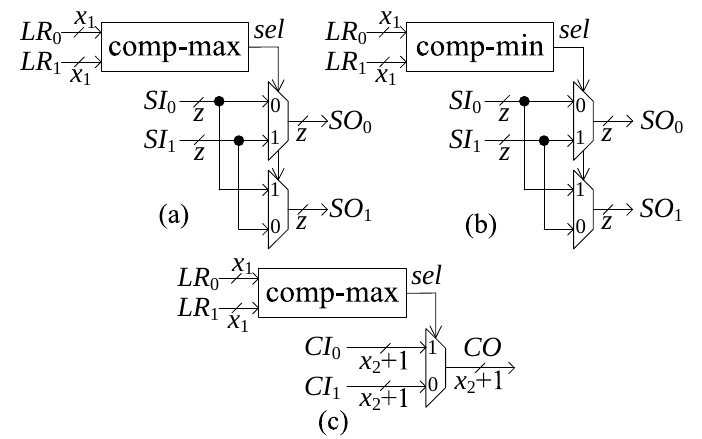}
  \caption{(a) Architectures of IS (b) Architectures of DS (c) Architectures of CAS ($z=x_1+x_2+1$)}\label{fig: is_ds}
\end{figure}

Each stage of the BSG consists of $\frac{L}{2}$ increase-order sorters (ISs) and $\frac{L}{2}$ decrease-order sorters (DSs), which are shown in Fig.~\ref{fig: is_ds}(a) and Fig.~\ref{fig: is_ds}(b), respectively. Both the IS and DS have two inputs and two outputs. For $k=0,1$, $SI_k=(LR_k, l_k, b_k)$ and $LR_k$, $l_k$ and $b_k$ denote the path metric and its corresponding list index and decoded bit. Besides, $SO_k=(LR_k', l_k', b_k')$ for $k=0,1$. The IS reorders the inputs such that path metric $LR_0' \leq LR_1'$. The output of the comp-max module is 1 when $LR_0 > LR_1$. The DS reorders the inputs such that $LR_0' \geq LR_1'$ and the output of the comp-min module is 1 when $LR_0 < LR_1$.

The BSG reorders the inputs based on the magnitude of path metrics. Let $LS_r (r=0,1,\cdots,15)$ denotes the associated path metric of output $S_r$, the path metrics of the 16 outputs satisfy:
\begin{eqnarray}
LS_0\leq LS_1 \leq \cdots \leq LS_7, \\
LS_8 \geq LS_9 \geq \cdots \geq LS_{15}.
\end{eqnarray}
It is proved in~\cite{bitonic_sorter} that the 8 maximum values among $LS_i$'s are $\max(LS_r, LS_{8+r})$ for $r=0,1,\cdots,7$.
Hence, a stage of CAS modules is appended at the outputs of BSG shown in Fig.~\ref{fig: mvf}, where CSA$_r$ takes $S_r$ and $S_{r+8}$ as inputs. This stage of CAS modules produce the outputs $O_t = (a_l, c_l)$ for $l=0,1,\cdots,L-1$. As shown in Fig.~\ref{fig: is_ds}(c), the CAS module compares the path metrics of its two inputs and selects the corresponding list index and bit value whose associated path metric is larger.

The metric sorter in~\cite{tree_list_dec} has the same function as that of the proposed MVF. We compare the proposed bitonic sorter based MVF module with the metric sorter~\cite{tree_list_dec} in terms of area and critical path delay (CPD) under different list sizes. As shown in Table~\ref{tab:bitonic_cmp}, both modules are synthesized under the TSMC 90nm CMOS technology. The RTL files of the metric sorter are provided by the authors of~\cite{tree_list_dec}. As shown in Table~\ref{tab:bitonic_cmp}, the proposed MVF module is more suitable for large list sizes. For list size $L=2$ to 32, the proposed MVF achieves 8\% to 77\% area saving. The proposed MVF architecture achieves area saving because the comparator dominates the area for the metric sorter and the MVF modules. For list size $L$, the metric sorter needs $N_{MS}=L(2L-1)$ comparators, while the proposed MVF module needs $N_{MVF}=1+2+\cdots+\log_2L=\frac{L}{2}((\log_2L)^2+\log_2L+2)$ comparators. When $L$ is large, $N_{MS}/N_{MVF}\approx \frac{4L}{\log_2L}$. Clearly, our MVF module needs fewer comparators.

When $L=2, 4, 8$, compared with the metric sorter, the proposed MVF has longer CPD while achieving area saving. However, the longer delay for the MVF is inconsequential because it is not in the critical path for the decoder architecture when $L\leqslant 8$. When $L=16, 32$, the proposed MVF is better than the metric sorter in terms of both area and CPD. Thus, the proposed MVF is more suitable for large list sizes.

\subsubsection{Crossbar Control Signal Generator}
\label{sssec: ccsg}
Due to the lazy copy method~\cite{tree_list_dec}, when decoding path $l$ needs to be copied to decoding path $l'$, instead of copying LLMs from path $l$ to path $l'$, the index references ($\mathbf{r}_l = (r_l[n-1],\cdots,r_l[0])$ shown in Algorithm~\ref{algo: metricComp}) to LLMs of path $l$ are copied to path $l'$.
For decoding path $l$, when PUA$_l$ is computing updated LLMs in $P_{l,\lambda}$, the crossbar (CB) module shown in Fig.~\ref{fig: top_archi} selects input LLMs from decoding path $r_l[\lambda-1]$. The CB can be implemented with $L$-to-1 multiplexors.

\begin{figure} [hbt]
\centering
  \includegraphics[width=3in]{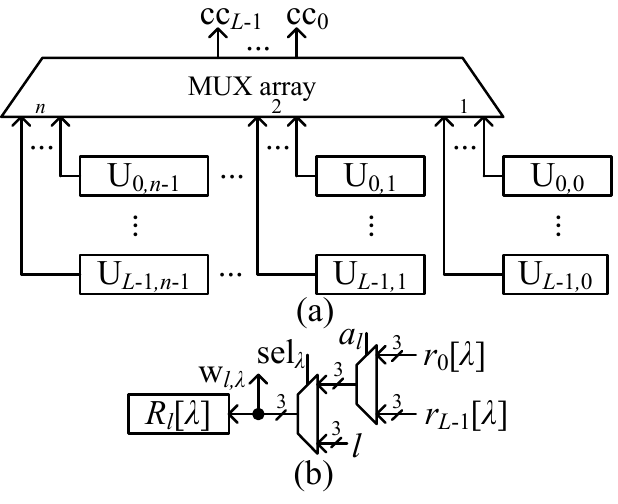}
  \caption{(a) The architecture of the proposed CCG (b) the architecture of a basic update unit}\label{fig: ccg}
\end{figure}

The crossbar control signal (CCG) generator computes the control signals of CB, cc$_0$, cc$_1$, $\cdots$, cc$_L-1$, where the $l$-th output of CB is connected to the cc$_l$-th input. An example of the CCG is shown in Fig.~\ref{fig: ccg}, where the proposed CCG consists of $Ln$ basic updating units, U$_{l,\lambda}$'s ($l=0,1,\cdots,L-1$ and $\lambda =0,1,\cdots,n-1$). As shown in Fig.~\ref{fig: ccg}(b), the proposed U$_{l,\lambda}$ contains an index register $R_{l}[\lambda]$ which stores $r_{l}[\lambda]$, where $r_l$ is the message updating reference index array for decoding path $l$.

When $u_i$ is being decoded, the multiplexors in Fig.~\ref{fig: ccg}(b) are configured so that w$_{l,\lambda}$ = $r_{a_l,\lambda}$ when $\lambda < \phi^{(i)}$ and w$_{l,\lambda}$ = $l$ otherwise. $\phi^{(i)}$ is defined in Section~\ref{ssec: ca_scl_algo}. When $P_{l,\lambda}$ needs to be computed, the $\lambda$-th inputs of the MUX in Fig.~\ref{fig: ccg}(a) are selected as the outputs of CCG. Once a round of metric computation is finished, w$_{l,\lambda}$ is written into its corresponding index registers.

\subsection{Partial Sum Update Unit and the CRC Unit}
\label{ssec:psu}
In this paper, a parallel partial sum update unit (PSU) is proposed to provide the partial sum inputs to $L$ PUAs when performing the $G$ computation. Compared with the PSU in~\cite{gross_polar1, tree_list_dec}, which needs $N-1$ single bit registers for a decoding path, our PSU needs only $\frac{N}{2}-1$ single register bits.

\begin{figure} [hbt]
\centering
  \includegraphics[width=3.4in]{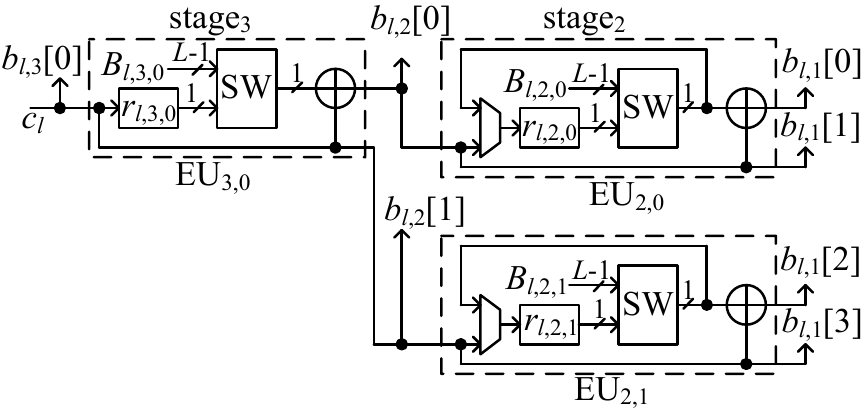}
  \caption{PSU architecture}\label{fig: psu}
\end{figure}

Take $N=2^3$ as an example, the architecture of PSU$_l$, which computes the partial sums for decoding path $l$, is shown in Fig.~\ref{fig: psu}, where stage$_3$ and stage$_2$ have one and two elementary update units (EUs), respectively. $r_{l,3,0}$, $r_{l,2,0}$, $r_{l,2,1}$ shown in Fig.~\ref{fig: psu} are single bit registers. $c_l=\hat{u}_{l,i}$ is the binary input of the PSU$_l$. There are three partial sum outputs: $b_{l,3}$, $b_{l,2}$ and $b_{l,1}$ with a width of 1, 2 and 4 bits, respectively. When the LLMs in $P_{l,\lambda}$ need to be updated with the $G$ computation, $b_{l,\lambda}$ is the corresponding partial sum input. The architectures of PSU$_l$ for other code lengths can be derived from the architecture in Fig.~\ref{fig: psu}. For a polar code with length $N=2^n$, the corresponding PSU$_l$ contains $n-1$ stages: stage$_n$, stage$_{n-1}$, $\cdots$, stage$_2$, where stage$_j$ has $2^{n-j}$ EUs for $n\geq j \geq 2$.

When bit index $i$ is even, $c_l$ is stored in $r_{l,n,0}$ and other registers keep their current values unchanged. When bit index $i$ is odd, bit registers in stage$_{n-1}$, stage$_{n-2}$, $\cdots$, stage$_{\phi^{(i+1)}-1}$ are updated with their corresponding input. When decoding path index $l \neq a_l$, the updated partial sums of decoding path $l$ should be computed based on the bit registers in PSU$_{a_l}$. The switch network (SW) shown in Fig.~\ref{fig: psu} selects the corresponding bit register value from PSU$_{a_l}$. The width of the input signal $B_{l,j,k} = \{r_{0,j,k}, r_{1,j,k}, \cdots, r_{L-1,j,k}\}\backslash \{r_{l,j,k}\}$ is $L-1$ bits.

\begin{figure*} [hbt]
\centering
  \includegraphics[width=6.6in]{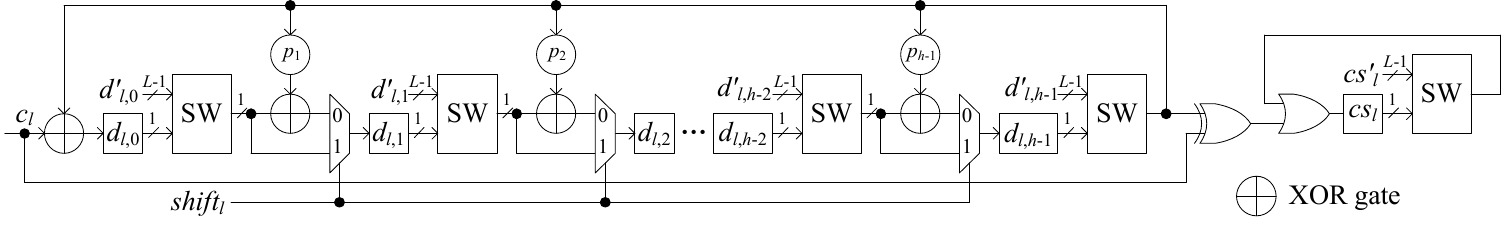}
  \caption{Architecture of the proposed CRC unit}\label{fig: crc}
\end{figure*}

The CRC unit (CRCU) checks whether a codeword passes the CRC. Suppose an $h$-bit CRC checksum is used, the architecture of the CRCU$_l$ for decoding path $l$ is shown in Fig.~\ref{fig: crc}, where the generation polynomial for the CRC checksum generation is $p(x) = x^h+p_{h-1}x^{h-1}+\cdots+p_1x+1$. The proposed CRCU$_l$ is based on a well known serial CRC computation architecture~\cite{serial_crc}. If the polynomial coefficient $p_k = 0$, the corresponding XOR gate and multiplexer are removed. During the decoding of the first $N-h$ code bits, the control signal \emph{shift}$_l=0$ and CRCU$_l$ computes the $h$-bit checksum of these code bits. The checksum is stored in bit registers $d_{l,0},d_{l,1},\cdots,d_{l,h-1}$ shown in Fig.~\ref{fig: crc}. Once the checksum computation is finished, the checksum is compared with the remaining $h$ decoded code bits and the control signal \emph{shift}$_l=1$. The checksum and the remaining $h$ code bits are compared bit by bit. The comparison result is stored in the register $cs_l$. The decoded codeword for decoding path $l$ passes the CRC only if \emph{cs}$_l = 0$. The SW module shown in Fig.~\ref{fig: crc} is the same as that used in the partial sum computation unit PSU$_l$. When $l \neq a_l$, the SW module selects $d_{a_l,k}$ for $k=0,1,\cdots,h-1$.

\begin{table*}[hbt]
  \centering
  \caption{Implementation Results With $R' = 0.468$ and $R=0.5$}
  \label{tab:imp_result}
  \begin{threeparttable}
  \footnotesize
  \begin{tabular}{c|c|c|c|c|c|c|c|c|c|c}
    \hline
     &\multicolumn{4}{c|}{proposed architecture} & \multicolumn{4}{c|}{\cite{tree_list_dec}\dag} & \multicolumn{2}{c}{\cite{tree_list_dec}\ddag}\\ \hline\hline
     algorithm & \multicolumn{4}{c|}{CA-SCL} & \multicolumn{6}{c}{SCL} \\ \hline
     list size $L$ & \multicolumn{2}{c|}{2} & \multicolumn{2}{c|}{4} & \multicolumn{2}{c|}{2} & \multicolumn{2}{c|}{4} & 2&4\\ \hline
     total number of PUs $LT$& 16& 32& 32& 64& 16& 32& 32& 64 & 128 & 256 \\ \hline
     channel message quantization bits $t$ & \multicolumn{8}{c|}{4} & \multicolumn{2}{c}{3}\\ \hline
     process & \multicolumn{8}{c|}{TSMC 90nm} & \multicolumn{2}{c}{UMC 90nm}\\ \hline
     frequency (MHz)    &500   & 500  & 454  & 476   &699    & 757   &684    & 694 & 459 & 314 \\ \hline
     total area (mm$^2$)&0.406 & 0.553& 0.810& 1.132 & 1.114 & 1.174 & 2.181 & 2.197 & 1.60 & 3.53\\ \hline
     $N_C$              & 3200 & 2816 & 3200 & 2816  & 3200  & 2816  & 3200  & 2816 & 2592 & 2592\\ \hline
     latency (ms)       & 6.4  & 5.63 & 7.04 & 5.91  & 4.57  & 3.71  & 4.67  & 4.05 & 5.64 &8.25 \\ \hline
     throughput (Mbps)  &160$R'$   & 181$R'$  & 145$R'$  & 173$R'$   &224$R$    & 275$R$   &219$R$    & 252$R$ & 181$R$ &124$R$  \\ \hline
     hardware efficiency (Mbps/mm$^2$)& 394$R'$ & 327$R'$ & 179$R'$ & 152$R'$ & 201$R$ & 234$R$ & 100$R$ & 114$R$ &113$R$ & 35$R$\\ \hline
     normalized hardware efficiency & 1.83 & 1.30 & 1.67 & 1.24 & 1 & 1 & 1& 1&\multicolumn{2}{c}{-}\\ \hline
  \end{tabular}
    \begin{tablenotes}[para,flushleft]
    \ddag Original synthesis results based on a UMC 90nm technology in~\cite{tree_list_dec}.

    \dag Synthesis results based a TSMC 90nm technology, provided by the authors of~\cite{tree_list_dec}.
  \end{tablenotes}
  \end{threeparttable}
\end{table*}

\subsection{Decoding Cycles}
\label{ssec:dec_latency}
For the proposed list decoder, pipeline registers can be inserted in the paths that pass through the MVF. 
Let $N_C$ denote the number of cycles spent on the decoding of one codeword. For list decoder architectures based on partial parallel processing~\cite{gross_polar1},
    \begin{equation}\label{equ: c_count}
    N_C=2N+\frac{N}{T}\log_2 \frac{N}{4T} + n_pRN,
    \end{equation}
where $N$, $T$, $n_p$, $R$ denote the block length, the number of PUs per decoding path, the number of pipeline registers inserted in the path pruning unit and the code rate, respectively.

The corresponding throughput $TP = \frac{fNR}{N_C}$, where $f$ is the frequency of the list decoder.
The latency $T_D = \frac{N_C}{f}$.


\subsection{Scalability of the Proposed List Decoder Architecture}
\label{ssec:dec_scalability}

Based on the FER results, our list decoder architecture is more suitable for list sizes since a larger $L$ leads to more performance gain for the CA-SCL algorithm. For the current list decoder architecture in~\cite{tree_list_dec}, there are two issues when $L$ increases.
\begin{itemize}
\item The message memories of the list decoder in~\cite{tree_list_dec} are built with registers due to the non-uniform quantization of the logarithm domain messages. Besides, the message memories dominate the whole decoder area. As a result, the memory area of the list decoder is linearly proportional to list size $L$. For a larger list size, the list decoder architecture in~\cite{tree_list_dec} will suffer from large area and high power consumption due to its register based memory.
\item As shown in Table~\ref{tab:bitonic_cmp}, when the list size grows, the metric sorter suffers from large area and long critical path delay, which results in a slower clock frequency of the list decoder. If multiple pipelines are inserted in the metric sorter, the number of cycles for decoding one codeword also increases as shown in Eq.~(\ref{equ: c_count}).
\end{itemize}
For our list decoder architecture, these two issues are solved as follows.
\begin{itemize}
\item The proposed memory architecture is more area efficient compared to register based memory. Besides, the proposed memory architecture offers a tradeoff between data throughput and memory area. The register based memory~\cite{tree_list_dec} remains almost unchanged when the number of PUs changes. However, for the proposed memory architecture, the number of PUs affects the depth-width ratio of the message memories. Hence, the area of message memory can be tuned by varying the number of PUs. Reducing the number of PUs will increase the depth of message memories, which is more area efficient. On the other hand, reducing the number of PUs will also increase the number of cycles used on decoding one codeword and decrease the data throughput.
\item When the list size increases, the proposed MVF is more area efficient and has a shorter critical path delay compared with the metric sorter~\cite{tree_list_dec}.
\end{itemize}

As shown in Eq.~(\ref{eqn: mem_depth}), the depth of the largest LLM memory instance will increase when $N=2^n$ increases. Hence, the area efficiency will be improved when $N$ increases. As a result, our list decoder architecture is more suitable for large block length $N$.

\section{Implementation Results}
\label{sec:result}
In this paper, our list decoder architecture has been implemented with list size $L$ = 2 and 4 for a rate 1/2 polar code with $N=1024$. For each list size, two list decoders with the numbers of $T$ = 8 and 16 PUs, respectively, are implemented and synthesized under a TSMC 90nm CMOS technology. For the L-MEM within each of our list decoder, each sub memory is compiled with a memory compiler if its depth is large enough. Otherwise, the sub memory is built with registers. For all implemented decoders, each channel LLM is quantized with 4 bits in order to achieve near floating point decoding performance. For our list decoders with $L=2$ and 4, one stage of pipeline registers is used. Since the synthesis results in~\cite{tree_list_dec} were based on a UMC 90nm technology, the authors of~\cite{tree_list_dec} have generously re-synthesized their decoder architecture using the TSMC 90nm technology. We list both synthesis results from~\cite{tree_list_dec} and the re-synthesized results provided by the authors of~\cite{tree_list_dec} in Table~\ref{tab:imp_result}. To make a fair comparison, we focus on the re-synthesized results.

Based on the implementation results in Table~\ref{tab:imp_result}, we have the following observations.
\begin{itemize}
\item The decoder architecture in~\cite{tree_list_dec} has higher a throughput than our list decoder architecture. The reason is that the decoder architecture in~\cite{tree_list_dec} employs register based memory while the proposed list decoder architecture employs register file (RF) based memories. The read and write delays of an RF are larger than those of a register based memory, respectively.
\item On the other hand, our list decoder architecture is more area efficient based on the area comparisons shown in Table~\ref{tab:imp_result}. In terms of the hardware efficiency, our list decoder architecture is better than that in~\cite{tree_list_dec}. For list decoders with the same $L$ and $T$ values, compared with the decoder of~\cite{tree_list_dec}, our list decoder architecture achieves 1.24 to 1.83 times of hardware efficiency.
\end{itemize}

Our list decoder is implemented for the $N=1024$ polar code because the same block length is used in~\cite{tree_list_dec}. For larger block length or larger list size, our advantage in hardware efficiency is expected to be greater due to more area efficient LLM memory.

Since the CA-SCL algorithm helps to select the correct one from $L$ possible decoded codewords~\cite{ido_list2}, the decoding performance of the CA-SCL algorithm is better than that of the SCL algorithm with the same list size in~\cite{tree_list_dec}.
As shown in Fig.~\ref{fig: fer}, the proposed CA-SCL decoders in Table~\ref{tab:imp_result} outperform the SCL decoders in Table~\ref{tab:imp_result}. We note that the number of PUs has no impact on the error performance of the SCL and CA-SCL decoders.

As shown in Fig.~\ref{fig: fer}, for the CA-SCL algorithm, increasing the list size results in noticeable decoding gain according to our simulations. As shown in~\cite[Fig.~1]{ido_list1}, increasing the list size of the SCL algorithm leads to negligible  decoding gain especially in high SNR region. For the CA-SCL algorithm, the choice of list size $L$ depends on the tradeoff between error performance and decoding complexity. Better error performance can be achieved by increasing the list size $L$. For the SCL algorithm, we need to find the threshold value $L_T$, where little further decoding gain is achieved by employing a list size $L>L_T$. For the SCL algorithm, the feasible list size should be no greater than $L_T$ and satisfy the error performance requirement.

Due to the serial nature of the successive cancelation method, the SC based decoders and its list variants suffer from long decoding latency. In terms of throughput, the throughput of SC based decoders is expected to be lower than BP based decoders, since the BP algorithm for polar codes has a much higher parallelism. On the other hand, the BP algorithm for polar codes still suffers from inferior finite length error performance~\cite{bp_polar_org, finite_length_polar}. Current simulation results~\cite{finite_length_polar} show that the error performance of the BP algorithm for polar codes is similar to that of SC algorithm, but worse than those of the SCL and CA-SCL algorithms.
\section{Conclusion}
\label{sec:conclusion}
In this paper, an efficient list decoder architecture has been proposed for polar codes. The proposed decoder architecture achieves higher hardware efficiency and better error performance than previous list decoder architectures. 

\section*{Acknowledgment}
We thank the authors of~\cite{tree_list_dec} for generating the synthesis results using the TSMC 90nm technology. We also want to express our gratitude to anonymous reviewers for their constructive comments.
\bibliographystyle{IEEEbib}
\bibliography{refs_latest}
\end{document}